\documentclass[10pt,conference]{IEEEtran}
\usepackage{cite}
\usepackage{amsmath,amssymb,amsfonts}
\usepackage{graphicx}
\usepackage{textcomp}
\usepackage{xcolor}
\usepackage[hyphens]{url}
\usepackage{fancyhdr}
\usepackage{hyperref}

\usepackage{algorithm}
\usepackage{xcolor}
\usepackage{makecell} 
\usepackage{graphicx}
\usepackage{enumerate}
\usepackage{subcaption} 
\usepackage{algpseudocode}
\usepackage{tcolorbox}
\usepackage{booktabs}  
\usepackage{multirow}
\usepackage{svg}
\usepackage{tabularx}
\usepackage{algpseudocode}
\usepackage{tikz}
\usepackage{hyperref}
\usepackage{pdfcomment}
\usepackage{cleveref}
\usepackage{balance}

\newcommand{\figgap}{6pt}          
\newcommand{\subfigheight}{3.0cm}  

\definecolor{darkyellow}{RGB}{204,153,0}

\newcommand{\revision}[1]{{\color{black}#1}}
\newcommand{\rebuttal}[1]{\textcolor{black}{#1}}
\newcommand{\clarify}[1]{\textcolor{black}{#1}}
\newcommand{\oioioi}[1]{\textcolor{black}{#1}}

\Crefname{equation}{Eq.}{Eqs.}
\Crefname{figure}{Fig.}{Figs.}
\Crefname{table}{Table}{Tables}
\Crefname{section}{Sec.}{Sec.}
\Crefname{subsection}{Sec.}{Sec.}

\newcommand{\fishicon}{%
  \raisebox{-0.3ex}{\includegraphics[height=0.9em]{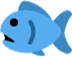}}%
}

\newcommand{\simplebox}[2]{%
\vspace{4pt}
\noindent
\fcolorbox{black}{cyan!65!blue!70!black!10}{%
  \begin{minipage}{0.97\columnwidth}
    \fishicon\hspace{0.2em}\textbf{#1}. #2
  \end{minipage}
}
}

\title{Robust \& Scalable Website Fingerprinting via Minimal DRAM Traffic}

\def\oioioicameraready{} 
\newcommand\oioioiauthors{Shubham Roy\textsuperscript{*},
Samar Perwez\textsuperscript{*},
and Biswabandan Panda}
\newcommand\oioioiaffiliation{Indian Institute of Technology Bombay, India}

\author{
  \ifdefined\oioioicameraready
    \IEEEauthorblockN{\oioioiauthors{}}
      \IEEEauthorblockA{
        \oioioiaffiliation{} \\
      }
  
  \fi 
}

\fancypagestyle{camerareadyfirstpage}{%

  \fancyfoot[C]{\thepage}
}

\begin{document}
\maketitle
\ifdefined\oioioicameraready
  \begingroup
    \renewcommand{\thefootnote}{*}
    \footnotetext{These authors contributed equally to this work.}
  \endgroup
  \setcounter{footnote}{0}
\fi

\ifdefined\oioioicameraready
  \thispagestyle{camerareadyfirstpage}
  \pagestyle{plain}
\else
  \thispagestyle{plain}
  \pagestyle{plain}
\fi




\newcommand{\oioioiheight}{0mm}
\ifdefined\eaopen
\renewcommand{\oioioiheight}{12mm}
\fi


\begin{abstract}
Website fingerprinting attacks infer which websites a victim visits and pose privacy concerns. Such attacks can be classified as on-path and co-located, based on the resources the attacker monitors. An on-path adversary, which runs on a different machine than the victim, observes all the network traffic of the victim. In contrast, a co-located attacker executes on the same machine as the victim and shares system resources with the victim. 
State-of-the-art co-located attacks, such as sweep-counting and loop-counting, leverage last-level cache (LLC) occupancy and interrupts, and achieve over 90\% classification accuracy for website fingerprinting. 
\clarify{However, the increasing prevalence of many-core processors with 16 cores and beyond in modern laptops, desktops, and workstations exposes a key limitation of existing attacks.} We show that loop-counting accuracy drops from 93\% on eight cores to 50\% on 24 cores as with an increase in core count, the number of interrupts handled per core decreases.
For the sweep-counting attack, the accuracy drops from 84\% when a many-core chiplet-based system with 16 cores shares one Core Complex Die (CCD) to 45\% when all 16 cores share two CCDs. 

To circumvent the scalability issues, we propose a memory-chasing website fingerprinting attack that exploits DRAM traffic interference \clarify{coming from websites}. Memory-chasing performs pointer-chasing traversals, which create accesses to DRAM. \clarify{Our attack with minimal interference captures the website's memory traffic, as the pointer-chasing access pattern of the attacker ensures at most one in-flight DRAM request at any given time.} 
The attacker counts DRAM accesses within fixed time periods to construct distinct traces for website classification. 
We evaluate our attack on the top-50 most-visited websites. In a closed-world setup, our attack achieves 96\% accuracy on a 24-core system, maintains 86\% accuracy across multiple CCDs, and demonstrates 80\% accuracy even on a system with multiple sockets. Additionally, our attack has an accuracy of 97\% in an open-world setup. 
Our evaluation across \clarify{many cores, LLC designs, multiple sockets, DRAM configurations, browsers, OSes, isolation mechanisms, system noise, and countermeasures} establishes memory-chasing as a robust and scalable attack.
\end{abstract}

\section{Introduction}\label{sec:introduction}
Co-located website fingerprinting (WF) attacks, by a remote attacker, exploit side channels, such as last-level cache (LLC)~\cite{shusterman2019robust} and system interrupts~\cite{jackcook}. These attacks are different from conventional WF attacks that exploit the network traffic patterns to predict the website visited by the victim~\cite{cai2012touching, hintz2002fingerprinting, herrmann2009website, hayes2016k}. In the co-located attack, a remote attacker embeds malicious JavaScript (JS) code within a compromised website to achieve co-location, allowing the JS code to co-run on a machine shared by a victim. To achieve reliable co-location, the remote attacker exploits phishing and malicious advertising~\cite{shusterman2019robust}. Highly accurate WF leads to privacy erosion and is also used for censorship and surveillance. Beyond its direct privacy impact, WF also enables subsequent attacks such as inter-keystroke timing inference, potentially reconstructing victim input~\cite{keytar}. Prior co-located WF attacks exploit a range of microarchitectural observables. SegScope~\cite{zhang2024segscope} captures interrupts induced by network and graphics activity during page loads. AVXProbe~\cite{kim2025avxprobe} leverages unprivileged AVX masked load/store instructions to probe TLB and cache regions of kernel modules responsible for rendering, networking, and cryptography. The \texttt{mwait} attack~\cite{ruiyi2023mwait} classifies websites by counting CPU wake-ups from system interrupts within a fixed time window. However, SegScope, AVXprobe, and mwait attacks require native code execution, unavailable to co-located JS-based attackers inside the browser sandbox. We therefore exclude these attacks due to their distinct threat model.

\begin{figure}[t]
  \centering
  \resizebox{\columnwidth}{!}{%
    \begin{tabular}{@{}c@{\hspace{\figgap}}c@{\hspace{\figgap}}c@{}}
      \includegraphics[height=\subfigheight]{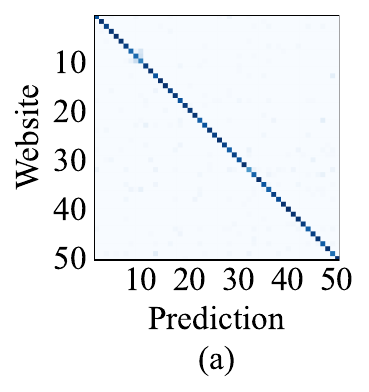} &
      \includegraphics[height=\subfigheight]{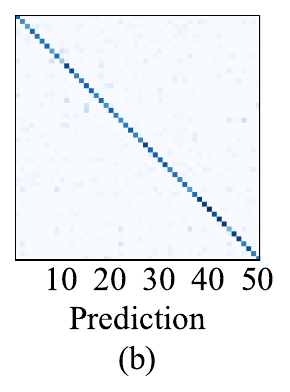} &
      \includegraphics[height=\subfigheight]{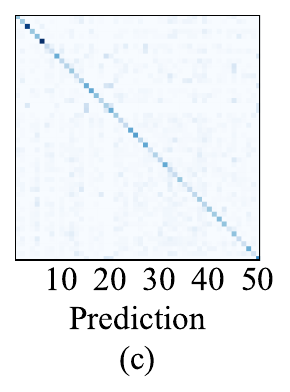}
    \end{tabular}%
  }
  \caption{WF confusion matrix of top-50 websites~\cite{siteList} with loop-counting~\cite{jackcook} on Intel Xeon Gold 6342 processor on (a) eight, (b) 16, and (c) 24 cores with respective classification accuracies of 93\%, 70\%, and 50\%.}
  \label{fig:confusion_matrix_loop}
  \vspace{-0.2in}
\end{figure}

Feng et al.~\cite{FishChips} present an evaluation of leakage sources of WF attacks. The authors compare sweep-counting~\cite{shusterman2019robust}, loop-counting~\cite{jackcook}, and mwait-based attacks~\cite{ruiyi2023mwait}, but do not address their scalability limitations. Note that among the three attacks, only loop-counting and sweep-counting deliver high classification accuracy ($>$90\%). In the sweep-counting attack, the attacker monitors the LLC occupancy when the victim loads a website by counting LLC sweeps possible in a fixed time period. Cook et al.~\cite{jackcook} argue that \emph{system interrupts} are the primary side channel and not LLC occupancy, even in the sweep-counting attack. The authors propose a loop-counting attack that classifies websites without any explicit memory accesses and argue that such accesses inhibit performance.

\textbf{The problem.} Modern desktop~\cite{amd9950x3d, intel7945hx, intel7980hx, intel10980xe} and laptop~\cite{intel7945hx} processors increasingly ship with 16 or more cores. \clarify{An exhaustive list of all the laptops and desktop processors with a core count greater than or equal to 16 cores is available here~\cite{cpu_config_0D86}.} \clarify{We exclude mobile processors, as none with 16 or more cores are commercially available yet.} Existing attacks, however, have only been demonstrated on systems with few cores. For instance, the loop-counting attack achieves a classification accuracy of 96\% on a multi-core system (Intel i5-2500)~\cite{shusterman2019robust, jackcook}, but it has not been evaluated on many-core systems.

We vary the number of CPU cores recognized by the Linux kernel~\cite{sysfs} and observe that classification accuracy with the loop-counting attack drops from 93\% on 8 cores to 70\% on 16 cores and to 50\% on 24 cores. 
Fig.~\ref{fig:confusion_matrix_loop} shows a confusion matrix for WF done on the top-50 websites~\cite{siteList} using the loop-counting attack, on eight, 16, and 24 cores on a system with an Intel Xeon Gold 6342 processor with a maximum of 24 cores.
Next, we quantify the sweep-counting attack on a chiplet-based AMD EPYC 9754 processor that has a Core Complex Die (CCD) based LLC. The accuracy of the sweep-counting attack drops from 84\% on 16 cores sharing one CCD to 45\% on 16 cores sharing two CCDs. 

\simplebox{Research questions}{Do state-of-the-art loop counting and sweep-counting attacks scale? If not, then why not?}
 


\textbf{Why is loop-counting not scalable?}
 The loop-counting attack captures the effect of system interrupts, generated when a website loads. Moreover, interrupts such as \emph{non-movable} \texttt{softirqs} play an important role in information leakage as compared to other movable interrupts like device interrupts or network \texttt{softirqs}, which can be isolated to a particular core~\cite{jackcook}.
 The adversary operates on a specific CPU core and observes only the \texttt{softirqs} scheduled to that core. In a many-core system, as \texttt{softirqs} can get handled by any core, the number of \texttt{softirqs} handled on a specific core decreases, reducing the attacker's observability.
 
  \simplebox{Observation I}{Loop-counting remains effective on systems with few cores, but accuracy drops as core count increases due to interrupts getting distributed across cores, reducing the attacker's observability.}


\textbf{Is sweep-counting scalable?}
To our surprise, the sweep-counting attack is scalable with an increasing number of cores, and its classification accuracy remains stable. As LLC is a shared resource, the effect of \emph{system interrupts} is still visible through LLC occupancy. \emph{Please note that the effectiveness of sweep-counting remains intact, not because of the LLC sweep that it performs, but because of the LLC activity generated from the system interrupts.} 
As the number of cores increases, the LLC size also increases accordingly. Different commercial processors use different design choices to accommodate this increase in LLC size. Intel has continued with a single logically monolithic LLC with LLC slices. However, AMD processors implement a CCD-based design, where each CCD contains Core Complexes (CCXs)~\cite{amd_doc} with each CCX consisting of a group of cores sharing a common LLC. Such CCD-based designs are also seen in desktop processors~\cite{isca_ccd}.

To assess the impact of CCDs, we fix the total number of cores to 16 and vary their placement across CCDs. When all 16 cores share a single CCD, sweep-counting achieves an accuracy of 84\%. However, distributing the cores across two CCDs reduces accuracy to 45\%, demonstrating the influence of LLC designs on the effectiveness of the sweep-counting attack.
\emph{We attribute this drop in accuracy to the limited cross-CCD influence of interrupts, as interrupts handled by cores in one CCD have limited interference with the LLCs in other CCDs}, thereby preventing the sweep-counting attack from detecting LLC activity because of interrupts.

\simplebox{Observation II}{Sweep-counting attack is susceptible to CCDs as its accuracy drops because of the distribution of system interrupts across CCDs.}

In summary, loop-counting and sweep-counting attacks observe the same underlying phenomenon, system interrupts, but through resources that do not scale. Loop-counting monitors the instruction throughput of a particular core, which is affected by interrupts handled by that core, but suffers as interrupts are distributed across cores. Sweep-counting monitors LLC occupancy but suffers from CCD-based designs. 

\textbf{Our goal} is to capture the \emph{big fish} (interrupts) while remaining scalable and robust to various system configurations. 

\textbf{How to capture the big fish?} We find that the DRAM traffic interference can be used for scalable WF, as DRAM is shared by all cores, even on a system with multiple sockets. Interrupt handling, memory accesses, and other system activity all manifest as DRAM traffic. We propose \emph{memory-chasing} attack, a scalable and robust way to perform WF. Our attack monitors DRAM traffic using a \emph{pointer-chasing} access pattern that ensures exactly one in-flight memory request, while still capturing all movable and non-movable interrupts across the system, irrespective of the core or socket the attacker resides in.
We count DRAM accesses within a given time period, which varies based on the DRAM traffic triggered by interrupts during a website load. We show that our attack is robust, and it can fingerprint websites even within a VM, establishing DRAM traffic as a reliable side channel for robust, scalable, and minimally intrusive website fingerprinting.

\textbf{Is it a DRAM timing channel attack?} No, as the memory-chasing attack does not time each access, like DRAMA~\cite{DRAMA}, using \texttt{rdtscp}, and is not intentionally introducing bank conflicts and row buffer conflicts to fingerprint websites.

\begin{figure}[t]              
    \centering
    \includegraphics[scale=0.5]{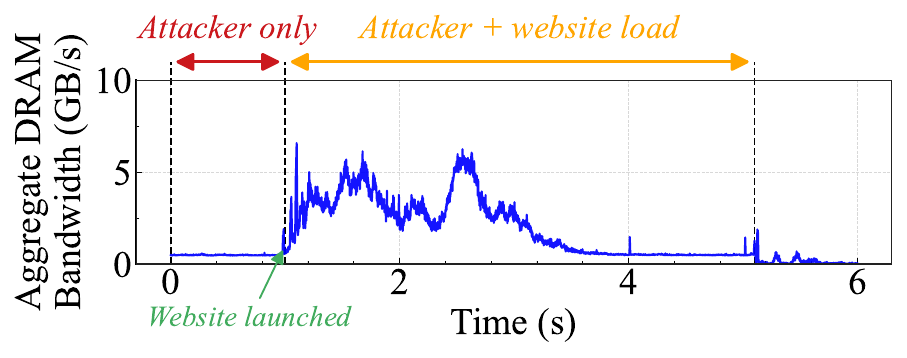}
    \caption{\clarify{DRAM bandwidth averaged over 100 runs while \texttt{chatgpt.com} loads. The attacker generates minimal memory traffic with an average bandwidth consumption of 0.5GB/s and RQ occupancy of 0.38 entries; the rise in bandwidth reflects webpage load, with negligible attacker bandwidth.}}
    \label{fig:bw_occ_plot}
    \vspace{-0.2in}
\end{figure}

\textbf{Is it a Denial of Service (DoS) attack?} The memory-chasing attack is not a DoS attack, unlike ~\cite{mutlu2017atc} and ~\cite{mc3date25} on DRAM bandwidth or Read Queue (RQ) of the DRAM controller. Fig.~\ref{fig:bw_occ_plot} profiles the memory-chasing attack on an Intel Xeon Gold 6342 processor with 8×16 GB DDR4 memory, using uncore performance monitoring and programmable hardware performance counters~\cite{perfdoc}. DDR4-2933 memory subsystem delivers 25.6 GB/s peak bandwidth per channel with a RQ capacity of 44 entries per DRAM controller. 
On average, the attacker occupies fewer than 0.38 RQ entries per DRAM cycle and consumes under 0.5 GB/s of DRAM bandwidth.
Pointer chasing serializes the attacker's requests by construction; each access depends on the completion of the previous access, bounding the attacker's occupancy to at most one in-flight request in memory.
Note that a typical website load consumes significantly higher DRAM bandwidth (5 GB/s) and RQ entries (4) than the attack itself. The attack, causes negligible contention at the RQ and DRAM bandwidth.

\textbf{Key results.} 
 For a closed-world setup, in which the attacker has complete knowledge of the websites visited by the victim, memory-chasing achieves 96\% accuracy on 24-core systems, 86\% across dual-CCD configurations, and 80\% in a dual-socket system. The attack remains effective with VM isolation with an accuracy of 78\% and works consistently across browsers and operating systems, including Tor browser with an accuracy of 91\%. Evaluation in an open-world setup, in which the attacker lacks complete knowledge of the websites visited by the victim, results in an accuracy of 97\%. 

\textbf{Our contributions.}  
(i) We analyze the scalability of loop-counting and sweep-counting attacks on many-core systems with different LLC designs. (Sec.~\ref{sec:loop_increase_core_count}).
(ii) We propose a memory-chasing WF attack, which performs pointer chasing to ensure DRAM accesses with an LLC-miss rate near 100\%. \clarify{By serializing accesses to a single in-flight request, memory-chasing generates negligible DRAM traffic, exploiting the interference that website loads already produce.} We demonstrate that counting individual DRAM accesses within fixed time periods captures DRAM traffic caused by website loads through system interrupts. \rebuttal{We show a scalable DRAM-based observation mechanism for known interrupt-driven leakage. }(Sec.~\ref{sec:the_right_fish}). 
(iii) We evaluate across varying core counts, LLC designs, multiple sockets, browsers, operating systems, and isolation mechanisms, establishing DRAM traffic as a robust side channel for scalable fingerprinting (Sec.~\ref{sec:eval}). 
(iv) We evaluate our attack with countermeasures proposed by Cook et al.~\cite{jackcook} and Shusterman et al.~\cite{shusterman2019robust} (Sec.~\ref{sec:countermeasure}).

\section{Background} \label{sec:background}
\subsection{Side-Channel Attacks}
Side-channel attacks exploit unintended information leakage. 
Recently, microarchitectural side channels have emerged as a threat, exploiting microarchitecture components.
\textbf{LLC-based side channels.} LLC attacks can be broadly classified as fine-grained timing or coarse-grained throughput attacks. Fine-grained attacks such as Prime+Probe~\cite{liu2015llc} and Flush+Reload~\cite{yarom2014flush} infer victim accesses by timing specific LLC sets or cache lines and typically require nanosecond-resolution timers. Coarse-grained attacks such as sweep-counting~\cite{shusterman2019robust} instead measure aggregate changes in LLC state over fixed intervals.

\textbf{Interrupt-based side channels.} Interrupts suspend execution and invoke kernel handlers for events such as network or timer activity. Linux splits handling into a time-critical top half and a deferred bottom half. \texttt{Softirqs} are high-performance deferred handlers that can execute concurrently on multiple CPUs and are commonly used for network packet processing~\cite{bovet2005understanding}. Interrupt-based side channels exploit the correlation between victim activity and the resulting interrupt frequency or timing.

\subsection{Website Fingerprinting (WF) Attack}

\begin{figure*}[t]
  \captionsetup[subfloat]{labelformat=empty} 
  \centering

  \setlength{\tabcolsep}{0.3em} 

  \begin{tabular}{ccc}
    \subfloat[(i)]{%
      \includegraphics[height=3cm,keepaspectratio]{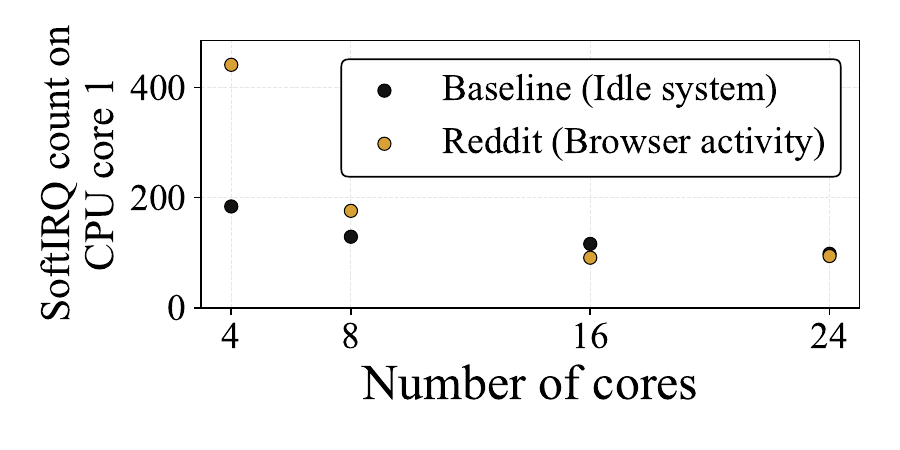}%
    } &
    \subfloat[(ii)]{%
      \includegraphics[height=3cm,keepaspectratio]{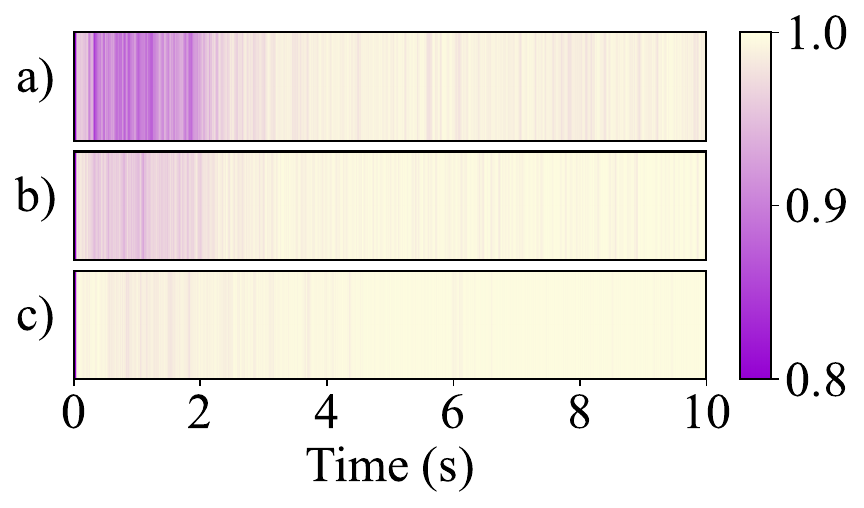}%
    } &
    \subfloat[(iii)]{%
      \includegraphics[height=3cm,keepaspectratio]{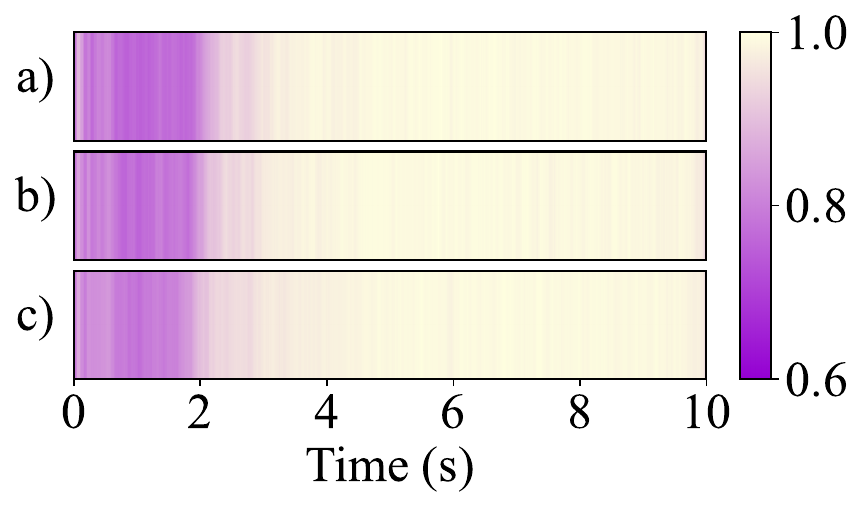}%
    } \\
  \end{tabular}

  \caption{(i) Per-core \texttt{softirq} count with increase in cores on System A (Table~\ref{tab:setup}). Counter values normalized to the maximum recorded with (ii) loop and (iii) sweep-counting for \texttt{reddit.com} averaged over 100 runs on a) eight, b) 16, and c) 24 cores.}
  \label{fig:combined_three}
\end{figure*}

Website fingerprinting attacks infer the websites visited by a victim, potentially revealing sensitive information about the victim's interests, activities, and behavior. Based on attacker placement, WF attacks are broadly classified as on-path or co-located. An on-path attacker observes traffic entering and leaving the victim's network; we adopt the threat model of Shusterman et al.~\cite{shusterman2019robust}. Our focus is on \emph{co-located attackers}, which execute on the same machine as the victim and share system resources. Since sandboxing and process isolation prevent direct observation of browsing activity, the attacker instead exploits shared-resource side channels to classify the visited website.

\textbf{Sweep-counting and loop-counting attack.} Shusterman et al.~\cite{shusterman2019robust,shusterman2021prime} propose the sweep-counting attack, which fingerprints websites by sampling LLC occupancy using an LLC-sized buffer over time periods of $P$ milliseconds. The attacker increments a counter (line~8) and performs an LLC sweep (line~9--11, Algorithm~\ref{algo:sweep}); as the victim website loads, it evicts the attacker's buffer entries, resulting in lower counter values that form a classifiable trace (line~13). Cook et al.~\cite{jackcook} revisit this attack and propose loop-counting, which eliminates the explicit memory accesses of line~9--11, yet achieves superior classification accuracy. More significantly, they demonstrate that the primary source of leakage is not LLC occupancy but an \emph{interrupt side-channel}: the trace captures variations in interrupt preemptions rather than LLC activity.

\begin{table}[t]
\centering
\footnotesize
\caption{System configurations}
\label{tab:setup}
\setlength{\tabcolsep}{3pt}
\begin{tabularx}{\columnwidth}{
    @{}
    >{\raggedright\arraybackslash}p{0.3\columnwidth}
    >{\raggedright\arraybackslash}X
    >{\raggedright\arraybackslash}X
    @{}
}
\hline
\textbf{Component} & \textbf{System A} & \textbf{System B} \\
\hline
CPU & Intel Xeon Gold 6342 & AMD EPYC 9754 \\
Cores/Threads & 24\,C / 48\,T & 128\,C / 256\,T \\
Memory & 128\,GB DDR4 @ 2933\,MT/s & 512\,GB DDR5 @ 4800\,MT/s \\
\rebuttal{Memory controllers} & \rebuttal{4} & \rebuttal{12} \\
\hline
\end{tabularx}
\vspace{-0.2in}
\end{table}
\section{The Elusive Big Fish}\label{sec:finding_the_right_fish}
In this section, we demonstrate that attacks such as loop-counting and sweep-counting lose their effectiveness in capturing accurate traces for distinguishing websites on a many-core system with a large LLC. We provide insights into why these attacks fail to achieve what they achieve in a small number of cores. Finally, we use the insights gained to motivate DRAM traffic as a resource for a scalable WF attack.

\begin{algorithm}[t]
\footnotesize
\caption{Sweep-Counting Attack}
\label{algo:sweep}
\begin{algorithmic}[1]
\State \Comment{P: sampling period in ms, T: total time in seconds}
\State buffer[LLC\_SIZE] \Comment{Buffer of LLC size}
\State Trace[T $\times$ 1000]
\While{loop}
\State counter $\gets$ 0
\State t\_begin $\gets$ \texttt{time()}
\Repeat
\State counter $\gets$ counter $+$ 1
\For{i $=$ 0 $\to$ size $-$ 1}
\State tmp $\gets$ buffer[i $\times$ 64]
\EndFor
\Until{(\texttt{time()} $-$ t\_begin) $\geq$ P} 
\State Trace[t\_begin] $\gets$ counter
\EndWhile
\end{algorithmic}
\end{algorithm}

\textbf{Evaluated systems.} In a many-core system, both the attacker and the victim can run on any core. Table \ref{tab:setup} shows the system parameters used for the WF attack. We use System A (Intel Xeon) to test scalability with core count. We use System B (AMD EPYC) for showcasing the effect of CCDs. \clarify{Both systems have a different number of memory controllers.} For all experiments on both systems, we use Ubuntu 20.04 LTS and Chrome 142 unless otherwise stated.
For the loop-counting attack, we experiment with increasing the core count from eight to 24 cores. For the sweep-counting attack, we experiment with one and two CCDs while keeping the core count fixed. Hardware prefetchers are enabled in all experiments. We consider the top-50 most visited websites~\cite{siteList}.

\textbf{Capturing the trace.}
The attacker's JS code runs concurrently in a different tab. Immediately after starting the attack, we navigate to the victim's website and record the trace. For loop-counting, we collect 2,000 samples at 5\,ms time periods for a total duration of 10\,s. For sweep-counting, we sample at 50\,ms periods over the same duration, as LLC sweeps take longer for larger LLCs. \rebuttal{Traces are captured without clearing the browser cache to emulate realistic browsing behavior.}

\textbf{Classifier.}
We use a Long Short-Term Memory (LSTM) network for website classification, following the hyperparameters of Cook et al.~\cite{jackcook} for its effectiveness in capturing temporal patterns. To prevent overfitting, we apply 10-fold cross-validation, splitting the dataset into 10 equal parts. In each iteration, one fold serves as the test set, while the remaining data splits into 81\% for training and 9\% for validation. Training stops when validation accuracy declines. The test fold is excluded entirely from training and validation, ensuring unbiased evaluation, and the average accuracy across all folds is reported. Since web pages are dynamic and evolve over time, training data collected at one point may not accurately represent future page content, leading to classifier performance degradation, commonly known as concept drift. We address this through incremental retraining as described by Shusterman et al.~\cite{shusterman206class}, which substantially mitigates this degradation while requiring only $\sim$2\% of the data needed for full retraining.

\subsection{Classification accuracy with loop-counting}\label{sec:loop_increase_core_count}
The classification accuracy of loop-counting drops from 93\% on eight cores to 70\% and 50\% on 16 and 24 cores, respectively, which shows that there is a significant drop in accuracy with the increase in core count.

\textbf{Why does classification accuracy degrade?} 
We illustrate this behavior using \texttt{reddit.com} as an example, although the trend holds across other websites. Fig.~\ref{fig:combined_three} (i) shows the number of \texttt{softirqs} delivered (using \texttt{/proc/softirqs}) to the attacker core with an increasing number of cores. We evaluate two cases: (a) when no website is active, and (b) during the load of \texttt{reddit.com}.
\noindent The loop-counting attack relies on capturing \texttt{softirqs} generated as a website loads, constructing a trace from these events, and subsequently using it for WF. Assuming the adversary operates on CPU core ID 1, only the \texttt{softirqs} scheduled to this core are observed and used to build the trace consisting of counter values.

\texttt{softirqs} are distributed across all cores, the per-core count of observed \texttt{softirqs} decreases as the number of cores increases. Consequently, the number of \texttt{softirqs} observed on CPU core ID 1 during the load of \texttt{reddit.com} approaches a case when no website is active (Fig.~\ref{fig:combined_three} (i)). Fig.~\ref{fig:combined_three} (ii) shows loop-counting traces for three different core counts. We normalize the counter values by dividing each value by the maximum counter value observed. We see that with an increased core count, the values increase, indicating more CPU execution and less time in interrupt handling.
The convergence of the per-core \texttt{softirq} at the core that is used for capturing the trace (Fig.~\ref{fig:combined_three} (i)) shows that the trace is not \emph{distinguishable} for \texttt{reddit.com}, and this holds across websites. As a result, the ML model struggles to distinguish between different websites, leading to a drop in accuracy for the loop-counting attack. In contrast, sweep-counting does not face this challenge with increased core count (Fig.~\ref{fig:combined_three} (iii)). However, in the next section, we show a case where sweep-counting fails to deliver high classification accuracy.

\simplebox{Takeaway I}{The time spent handling interrupts on the attacker CPU decreases as the number of cores increases.}
\subsection{Classification accuracy with sweep-counting}\label{sec:sweep_counting_limit}
To evaluate the effect of CCD, we use an AMD EPYC 9754 processor (System B in Table~\ref{tab:setup}), comprising eight CCDs, each with two CCXs and 16MB LLC per CCX.
On the AMD processor, we activate 16 cores in two configurations: (a) all 16 cores within a single CCD (both CCXs enabled) and (b) 16 cores distributed across two CCDs with one CCX per CCD.

\textbf{Classification accuracy:}
On the AMD processor, accuracy drops from 84\% for one active CCD to 45\% for two CCDs.


    
    


\begin{figure}[t]              
    \centering
    \includegraphics[width=0.9\columnwidth]{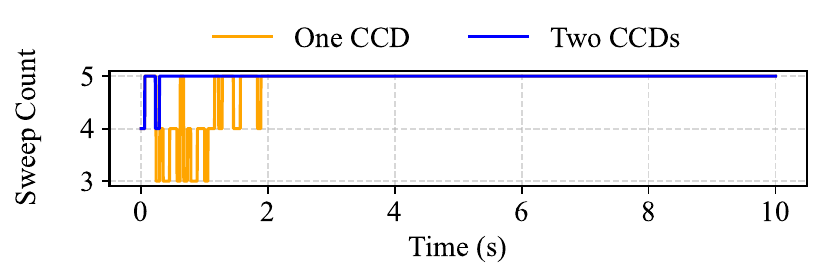}
\caption{Sweep-counting traces for \texttt{youtube.com} on System B (refer Table~\ref{tab:setup}) on 16 cores sharing one and two CCDs.}
    \label{fig:sweep_ccd_comparison}
    \vspace{-0.2in}
\end{figure}
\textbf{Why does classification accuracy drop?}
To handle an interrupt, the core transitions to interrupt context, after which the Interrupt Service Routine (ISR) executes. Once the ISR completes, the kernel may process any pending \texttt{softirqs} before finally returning to the interrupted context (either user mode or kernel mode). Loop-counting captures the lost CPU time due to interrupts on its core; however, sweep-counting captures LLC activity due to interrupts on cores sharing the same LLC.
\emph{The memory access in line no.~9 of Algorithm~\ref{algo:sweep} allows sweep counting to capture LLC effects of interrupts.} 

For CCD-based designs, the LLC is distributed across all the CCDs, and AMD uses Infinity Fabric to connect all CCDs through an I/O die~\cite{amd_doc}. Fig.~\ref{fig:sweep_ccd_comparison} compares the sweep-counting trace, recorded with cores sharing one CCD versus cores sharing two CCDs. We observe that for traces with two CCDs, the effect of interrupts is limited as interrupts executing on a given CCD have limited effects on LLCs in other CCDs. This is evident from Fig.~\ref{fig:sweep_ccd_comparison}, which illustrates a single trace of \texttt{youtube.com}. The sweep-counting trace with cores sharing two CCDs records lesser variation in sweep count compared to the trace where all cores share one CCD, making the trace appear flatter. These observations hold for other websites also.

\simplebox{Takeaway II}{On CCD-based CPUs, accuracy of sweep-counting decreases as the attacker has limited observability of cache effects when interrupts are handled on another CCD.}


\oioioi{\textbf{\hypertarget{rev:serial}{Effect of serialization and access pattern}.} The default sweep-counting attack uses randomized LLC accesses to avoid triggering hardware prefetchers~\cite{jackcook}. Replacing the randomized pattern with sequential accesses yields a similar classification accuracy of 43\%, compared to 45\% with randomized accesses, indicating that hardware prefetching has only a marginal effect on the attack. In contrast, replacing parallel accesses with dependent, serialized accesses reduces classification accuracy from 45\% to 26\%. These results show that sweep-counting benefits primarily from memory-level parallelism (MLP), which accelerates LLC fills and enables more sweeps within the observation window, rather than from avoiding hardware prefetching.}

\subsection{Why not have a \oioioi{collaborative} attack?} \label{subsec:multithread}
\oioioi{\textbf{Collaborative multi-process attack.} Both sweep-counting and loop-counting are JS-based attacks. One way to scale these attacks on many-core systems is to run multiple attacker processes concurrently and combine their traces. In a browser, however, the attacker \emph{cannot} directly spawn processes (websites). As browsers isolate websites in separate processes, multiple websites are needed with synchronization across them. A more practical alternative is to spawn multiple threads.} 

{\textbf{Collaborative multi-threaded attack.} An intuitive way to scale state-of-the-art attacks is to use multi-threading to collectively monitor the effect across all the cores. To monitor CPU throughput or LLC occupancy collectively, we need shared variables accessible across threads. The \texttt{SharedArrayBuffer} object in JS provides this functionality. 
However, this object is disabled in privacy-preserving browsers like Tor~\cite{torgit}~\cite{rokicki2021sok} and is restricted in other browsers like Chrome, Firefox, and Safari to cross-origin isolated contexts~\cite{coop}. We relax the practical constraints and use shared variables to evaluate multi-threaded variants of both loop and sweep-counting. Our evaluation shows that sweep-counting does not scale effectively due to contention among threads over the LLC-sized buffer. In contrast, while multi-threaded loop-counting does scale, its viability is constrained by the strict shared-memory requirements.

\textbf{Capturing the big fish.} Given the limitations that we presented in Sections~\ref{sec:loop_increase_core_count} and~\ref{sec:sweep_counting_limit}, we need a new attack that is robust and scalable. To capture ``the big fish" (system interrupts), we go one level deeper in the memory hierarchy to shared DRAM. In the next section, we present our \emph{memory-chasing} attack, which uses DRAM traffic for WF.

\section{Casting a Net for the Big Fish}
\label{sec:the_right_fish}
\textbf{Threat model.} Our threat model follows Cook et al.~\cite{jackcook} and Shusterman et al.~\cite{shusterman2019robust}. An attacker co-located with a victim executes on the same physical host and leverages shared microarchitectural resources. While sandboxing and process isolation prevent direct observation of the victim's browser session, the attacker can monitor shared resources to infer the victim’s activity. With limited privileges, the attacker can achieve co-location by tricking the victim into loading a malicious website that runs attacker-controlled JS code. The attacker is limited to the low-resolution timer provided by the browser. While a sensitive webpage loads, the attacker monitors the shared resources to classify websites. Realistic ways to achieve co-location even if the browser isolates memory, scheduling, and timer APIs, include (i) phishing: the user is tricked into visiting an attacker-controlled site containing malicious JS code, and (ii) malicious advertising: the attacker pays an advertisement network to serve malicious JS code through legitimate third-party sites~\cite{shusterman2019robust}.

\textbf{Attack overview.} We propose \emph{memory-chasing} attack, that exploits DRAM traffic rather than LLC occupancy. Memory-chasing performs pointer-chasing traversals on a buffer $k$ times larger than the LLC, ensuring that memory accesses bypass caches and reach off-chip DRAM, enabling the attacker to observe DRAM traffic with minimal interference. \emph{Our key insight is that DRAM traffic reflects system-wide activity, including interrupt handling and victim DRAM traffic, with visibility across all cores sharing the DRAM.} 
In the memory-chasing attack, we capture the number of DRAM accesses for a fixed time period, and construct a trace that fingerprints websites based on their DRAM traffic over time. The large resident set size of browsers, typically in hundreds of MiBs, ensures that browser activity generates observable DRAM traffic. A smaller DRAM access count within a period indicates increased DRAM activity from co-running applications (e.g., a website load), whereas a higher count implies lesser traffic at DRAM. By monitoring DRAM accesses, we capture the effect of both movable and non-movable interrupts, even if movable interrupts are pinned to a specific core.

\textbf{Attack description.} Algorithm~\ref{algo:memoryChasing} shows the memory-chasing attack. The attacker allocates a large buffer (\texttt{buffer}) of size \texttt{k} $\times$ LLC and builds an auxiliary array \texttt{next} that contains a randomized permutation of the \texttt{buffer} indices. For a fixed interval of \texttt{P} ms, the attacker continuously increments a counter (line no. 11), reads the next index from \texttt{next} array and performs a memory access to \texttt{buffer}, and uses the returned value in a small computation (line nos. 11–12) to prevent compiler optimizations. The attacker updates \texttt{current\_node} from \texttt{next} so the sequence of indices stays irregular. The \texttt{next} array visits every \texttt{buffer} index exactly once in a random order; this defeats hardware prefetchers and forces the accesses to reach DRAM. \emph{Our key insight is that the pointer-chasing access pattern serializes memory accesses, bounding in-flight DRAM requests to one, generating the minimal traffic necessary to capture DRAM activity.} After the interval of \texttt{P} ms, the attacker stores the counter in the \texttt{Trace} array (line no. 15). Finally, we record the trace for \texttt{T} seconds based on the \texttt{counter} values obtained after each \texttt{P} ms interval. In practice, \texttt{T} of 10 seconds suffices because only the website’s initial behavior needs to be captured.

\begin{algorithm}[t]
\footnotesize

  \caption{Memory-Chasing Attack}
  \label{algo:memoryChasing}
  \begin{algorithmic}[1]
    \State \Comment{P: sampling period in ms, T: total time in seconds}
    
    \State buffer[k $\times$ LLC\_size] \Comment{Buffer $k$ times LLC capacity}
    \State next$\text{\raisebox{0pt}[1.5\baselineskip][1.2\baselineskip]{\ensuremath{\left[\dfrac{\text{k}\times\text{LLC\_size}}{\text{cache\_line\_size}}\right]}}}\;$ \Comment{Auxiliary  index array}
    \State Trace[T$\times 1000$]
    \State current\_node $\gets$ 0
    \While{true}
      \State counter $\gets$ 0
      \State t\_begin $\gets$ \texttt{time()}
      \State sum $\gets$ 0
      \Repeat
        \State counter $\gets$ counter + 1
        \State current\_node $\gets$ next[current\_node]
        \State sum $\gets$ sum + buffer[current\_node $\times$ \text{cache\_line\_size}]
      \Until ({\texttt{time()} $-$ t\_begin) $\geq$ P}
      \State Trace[t\_begin] $\gets$ counter
    \EndWhile
  \end{algorithmic}
\end{algorithm}

\textbf{Bypassing LLC for DRAM.} Next, we examine the parameter \texttt{k}, which controls the buffer size. The LLC miss rate increases sharply as the buffer grows beyond the LLC capacity: we observe a miss rate of 2\% for a buffer size of $0.5\times \text{LLC}$, 36\% for LLC, 77\% for $2\times \text{LLC}$, 96\% for $4\times \text{LLC}$, and 99\% for $8\times \text{LLC}$. Larger \texttt{k} drives the attack to primarily capture off-chip DRAM traffic by forcing accesses to DRAM. We use $\texttt{k}=4$. The pointer-chasing access pattern ensures that most accesses reach DRAM without thrashing the LLC. Note that our goal is not to evict the victim’s cache lines from the LLC, but to prevent our own buffer from fitting in the LLC. Modern browsers and websites have memory footprints on the order of hundreds of MBs, already exceeding LLC capacity. 



\subsection{\clarify{Capturing DRAM access patterns}} \label{subsec:mem_dram_band} 
In this section, we ask: \clarify{\emph{does memory-chasing effectively distinguish different memory access patterns?}} To demonstrate the effectiveness of capture, we consider two synthetic memory-intensive workloads with distinct access patterns: sequential and irregular. Both frequently access DRAM and operate on identical working sets of a few gigabytes (e.g., 3 GiB). After initializing memory, the sequential workload traverses memory sequentially, benefiting from memory-level parallelism (MLP) and prefetching. In contrast, the irregular workload performs pointer chasing serializing dependencies with zero MLP. 

Fig.~\ref{fig:synthetic_bench} illustrates traces observed by the memory-chasing attacker for these co-running workloads, showing that the attack can distinguish between the two workloads by tracking their DRAM traffic. As shown in Fig.~\ref{fig:synthetic_bench}(a), for the sequential workload, the attacker observes the counter values to be lower compared to the irregular workload as shown in Fig.~\ref{fig:synthetic_bench}(b). This is because sequential accesses exploit MLP, resulting in a lower counter value in the memory-chasing attack. However, the irregular workload does pointer chasing, resulting in less DRAM traffic and consequently a higher counter value. 
\begin{figure}[t]
  \centering
  \includegraphics[width=0.8\linewidth]{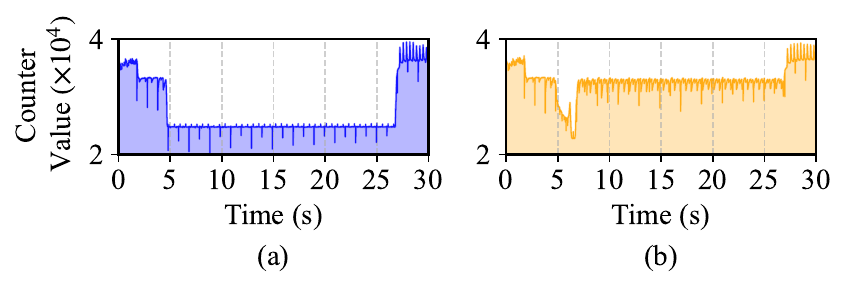}
  \caption{Counter values recorded at intervals of 5ms averaged over 15 runs captured with memory-chasing attack on System A (refer Table~\ref{tab:setup}) for two memory-intensive workloads with different access patterns: a) sequential and b) irregular.}
  \label{fig:synthetic_bench}
  \vspace{-0.2in}
\end{figure}


\begin{figure}[t]              
    \centering
    \includegraphics[width=0.8\columnwidth]{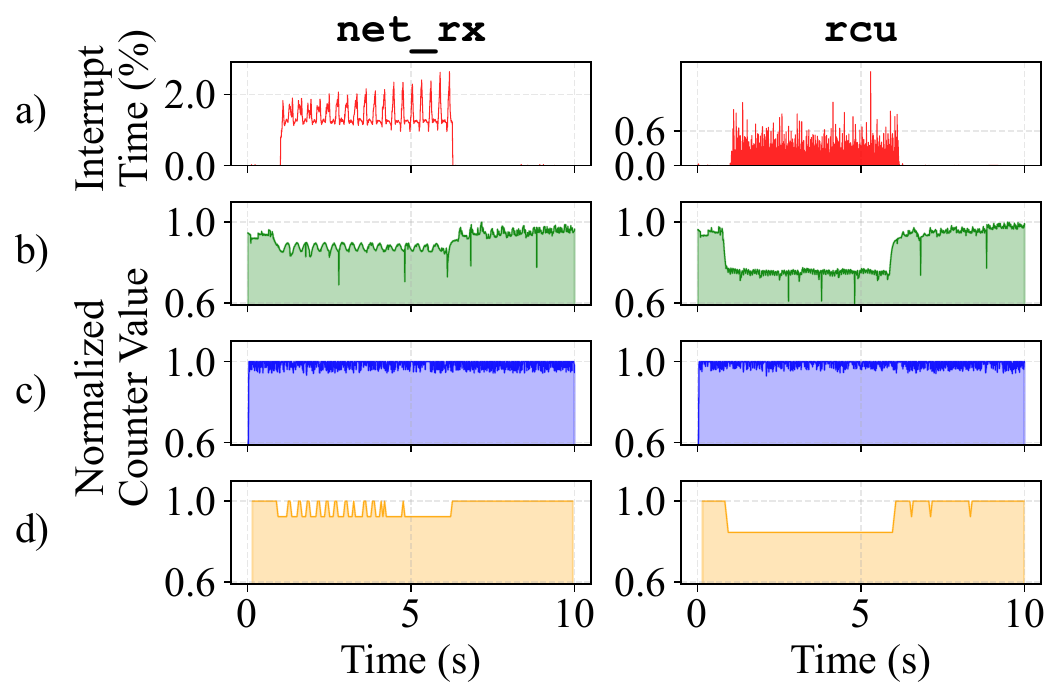}
    \caption{a) Percentage of CPU time spent processing \texttt{softirqs} averaged across all cores on System A (refer Table~\ref{tab:setup}) having 24 cores. Effect of synthetic workloads on max-normalized counter values recorded with b) memory-chasing, c) loop-counting, and d) sweep-counting attacks.}
    \label{fig:softirq_mc_loop_new}
    \vspace{-0.2in}
\end{figure}
\subsection{Capturing interrupts through DRAM traffic} \label{subsec:mem_dram_band} 
In this section, we ask, \emph{does off-chip DRAM traffic reveal fine-grained interrupts, such as \texttt{softirqs}?} To demonstrate that our attack captures \texttt{softirqs} at DRAM, we extend the eBPF code from Cook et al.~\cite{jackcook}. We track the \texttt{softirq} entry and exit on every core using Linux tracepoints \texttt{softirq\_entry} and \texttt{softirq\_exit}. We correlate these per-core \texttt{softirq} events with the memory-chasing traces.

We create synthetic workloads to trigger \texttt{net\_rx} and \texttt{rcu} \texttt{softirq} events while capturing corresponding traces for memory-chasing attacks. To evaluate cross-core visibility, all interrupts are injected on a core different from the attacker core. We use \texttt{net\_rx} and \texttt{rcu softirqs} to represent movable and non-movable interrupts, respectively, as \texttt{net\_rx softirqs} can be pinned to a specific core, unlike \texttt{rcu softirqs}. We trigger \texttt{net\_rx softirqs} through repeated \texttt{ping} requests. To induce frequent \texttt{rcu softirq} events, we concurrently execute numerous short-lived user processes on the target core, whose creation and termination trigger repeated read-copy-update (RCU) operations. 

We employ Linux kernel tracepoints to log \texttt{softirq} entry and exit events. We create epochs of 5\,ms and calculate the percentage of time each core spends servicing the targeted \texttt{softirq} during each epoch. We then compute the average of these percentages across all cores. Fig.~\ref{fig:softirq_mc_loop_new}(a) shows the percentage of CPU time spent handling \texttt{softirqs} averaged across all the cores. Fig.~\ref{fig:softirq_mc_loop_new}(b) shows the corresponding memory-chasing traces. As we can clearly see, the memory-chasing attack produces fine-grained traces that reliably capture \texttt{softirq} effects, including \texttt{net\_rx} interrupts, which are exclusively serviced on a different core, highlighting its robustness.  
However, loop-counting and sweep-counting do not effectively capture the interrupts as shown in Figs.~\ref{fig:softirq_mc_loop_new}(c) and (d) respectively. For a fair comparison across three attacks, we normalize the counter values by dividing them by the maximum counter value.


\simplebox{Takeaway III}{Memory-chasing attack captures DRAM traffic of \texttt{softirqs} distributed across all cores.}


\textbf{\rebuttal{Effect on victim's load time.}}
\rebuttal{We quantify any performance degradation on the victim's website load and evaluate the detectability of the memory-chasing attack. We use Largest Contentful Paint (LCP), which marks when the page's main content loads~\cite{lcp}. 
Each website is loaded 100 times under two scenarios: with and without the memory-chasing attacker. The average performance slowdown in terms of LCP is $<$2\% across the top-50 websites. We further quantify the difference using Cliff's Delta~\cite{cliff_delta}, which is positive and below 0.18 across all websites, indicating negligible performance overhead.}


\subsection{The illusion of correlation}\label{subsec:mem_loop_same}
In this section, we ask, \emph{are loop-counting and memory-chasing traces correlated?} To analyze traces of loop-counting and memory-chasing attacks, we collect data for three representative websites \texttt{chatgpt.com}, \texttt{x.com}, and \texttt{whatsapp.com} on an Intel system with a Xeon Gold 6342 processor (System A, Table~\ref{tab:setup}). As shown in Sec.~\ref{sec:loop_increase_core_count}, loop-counting is effective only with a few cores; hence, we record its traces using four cores, while memory-chasing traces are captured on four and 24 cores. On a four-core system, both attacks are expected to show high correlation. However, on a 24-core system, loop-counting does not capture \texttt{softirq} effects because \texttt{softirqs} are distributed across all cores. We therefore compare the 24-core memory-chasing trace with the four-core loop-counting trace. For fair comparison, we average counter values across 100 runs and normalize each by its maximum observed value. A strong correlation between these two attacks indicates that memory-chasing \emph{indeed} captures the effect of \texttt{softirqs}, unlike the myopic view of the loop-counting attack limited to the core in which the attacker runs. 

Fig.~\ref{fig:loop_mc_comparison}(a) shows that with four cores, loop-counting and memory-chasing attack traces are strongly correlated with the correlation coefficients of 0.87 and 0.91, for \texttt{chatgpt.com} and \texttt{x.com}. Fig.~\ref{fig:loop_mc_comparison}(b) shows that the correlation remains strong even for memory-chasing traces captured on 24-core, correlated with loop-counting traces captured on four cores, with correlation coefficients of 0.81 and 0.88 for \texttt{chatgpt.com} and \texttt{x.com}, respectively. Note that by recording traces on 24 cores, the \texttt{softirqs} are distributed across 24 cores. This reduces \texttt{softirqs} handled on the attacker's core, isolating DRAM traffic's contribution to the memory-chasing trace from CPU instruction throughput. Thus, we claim that the memory-chasing attack captures the same underlying event as the loop-counting attack, i.e., system interrupts, but via DRAM traffic. We showcase that although both loop-counting and memory-chasing traces correlate with each other, the \emph{cause} is not the same. Memory-chasing captures interrupts through DRAM traffic, whereas the loop-counting captures the effect through CPU instruction throughput.


\simplebox{Takeaway IV}{Memory-chasing attack on 24 cores captures the same trace as loop-counting attack on four cores by monitoring DRAM traffic.}

\simplebox{Summary}{Memory-chasing captures the interference at DRAM due to \texttt{softirq} activity during website loading. The pointer-chasing access pattern ensures only one in-flight request and generates minimal DRAM traffic.}
\begin{figure}[t]              
    \centering
    \includegraphics[width=0.9\columnwidth]{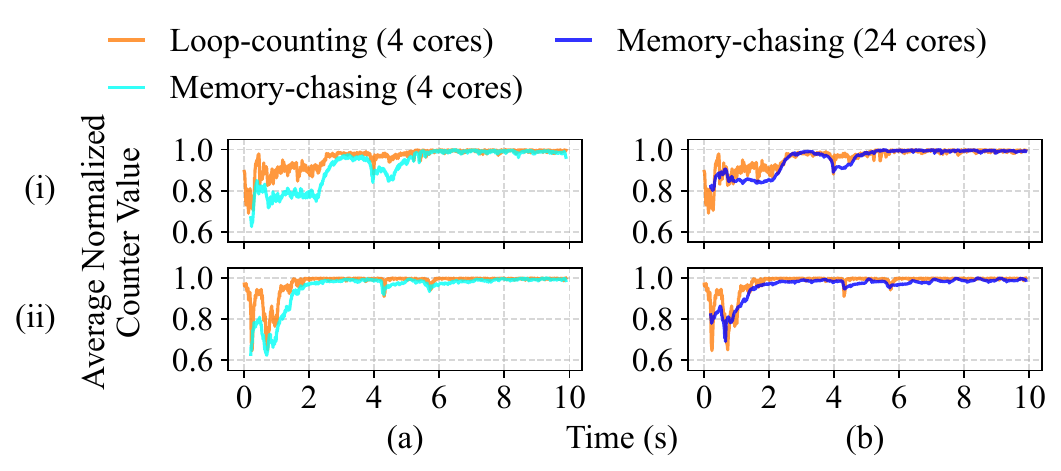}
    \caption{Normalized counter values, averaged over 100 runs on System A (refer Table~\ref{tab:setup}). Comparison of (a) loop-counting and memory-chasing traces with four cores and (b) loop-counting trace on four cores and memory-chasing trace on 24 cores for (i) \texttt{chatgpt.com} and (ii) \texttt{x.com}}
    \label{fig:loop_mc_comparison}
\end{figure}
\begin{table}[t]
\footnotesize
\centering
\caption{Classification accuracy (\%) for sweep-counting and memory-chasing across closed and open-world setups on System A (refer Table~\ref{tab:setup}) having 24 cores.}
\label{tab:open-close}
\setlength{\tabcolsep}{4pt}
\begin{tabular}{@{} l c ccc @{}}
\hline
\multirow{2}{*}{\textbf{Method}} &
\multirow{2}{*}{\textbf{\makecell{Closed-\\World}}} &
\multicolumn{3}{c}{\textbf{Open-World}} \\
\cmidrule(l){3-5}
 & & \textbf{Sensitive} & \textbf{\makecell{Non-\\Sensitive}} & \textbf{\makecell{Combined\\Accuracy}} \\
\hline
\textbf{Memory-chasing} & \textbf{96} & \textbf{96} & \textbf{100} & \textbf{97} \\
Sweep-counting          & 94          & 94          & 99           & 96 \\
Loop-counting           & 50          & 39          & 97           & 58 \\
\hline
\end{tabular}
\vspace{-0.1in}
\end{table}
\section{Evaluation}\label{sec:eval}
In this section, we comprehensively evaluate the memory-chasing attack. We demonstrate classification accuracy across various configurations, including different core counts and LLC designs. We assess the attacker's effectiveness first in a closed-world setup and then in an open-world setup. We also demonstrate its robustness across different browsers and operating systems. \rebuttal{Unless otherwise stated, we do not pin the attacker or victim to any CPU core}. Additionally, we evaluate classification accuracy with the attacker and victim running on separate VMs and on a system with multiple sockets. Finally, we assess the attack's robustness against countermeasures, including cache-sweep noise~\cite{shusterman2019robust}, spurious interrupt noise, and timer-based defenses proposed by Cook et al.~\cite{jackcook}.

\textbf{Closed-world and open-world.}
We evaluate attacks under two setups: closed-world and open-world. In the closed-world setup, the attacker knows all websites the victim visits and aims to distinguish among 50 websites selected from the top-50 most visited websites on Wikipedia~\cite{siteList}. \rebuttal{We treat the same website across different regional domains as distinct classes.} We collect 100 traces per website. In the open-world setup, the attacker lacks complete knowledge of visited websites. The victim accesses both ``sensitive" and ``non-sensitive" websites, with the attacker knowing only the sensitive set (i.e., websites in a closed-world setup) and aiming to identify accesses to these websites while labeling others as ``non-sensitive". For evaluation, we treat the top-50 closed-world websites as sensitive websites and collect 2500 additional traces from unique ``non-sensitive" websites in the Alexa top websites list, following Cook et al.~\cite{jackcook}. Unless otherwise stated, we evaluate on Chrome running on System A (Table~\ref{tab:setup}). We use the ML model described in Sec.~\ref{sec:loop_increase_core_count}, with traces of length 10 seconds for Chrome 142 and Firefox 136, and 30 seconds for Tor 15, as its security mechanisms slow page loads.

\textbf{Evaluated systems.} We conduct our experiments on two systems. We use System A (refer Table~\ref{tab:setup}), having 24 cores, as our primary system. For multiple socket experiments, we use a dual-socket system with two of these Intel CPUs. This configuration represents the most challenging attack scenario, where attacker and victim execute on two separate processors. For CCD-based evaluation, we use System B (refer Table~\ref{tab:setup}).

\subsection{Classification Accuracy}\label{subsec:accuracy}
\begin{figure}[t]              
    \centering
    \includegraphics[width=0.8\columnwidth]{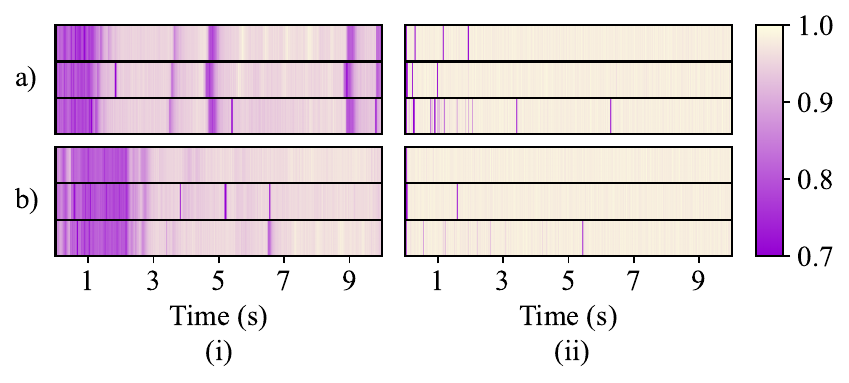}
    \caption{Counter values normalized to the maximum value recorded with (i) the memory-chasing and (ii) the loop-counting attacks for three runs of (a) \texttt{amazon.com} and (b) \texttt{booking.com} on System A (refer Table~\ref{tab:setup}) having 24 cores.}
    \label{fig:mc_vs_loop}
    \vspace{-0.2in}
\end{figure}
\textbf{Closed-world.}
In the closed-world setup, the memory-chasing attack demonstrates high classification accuracy regardless of the number of cores in the system. On eight cores, memory-chasing has a classification accuracy of 97\%. On 24 cores, the memory-chasing attack achieves a classification accuracy of 96\%, compared to 50\% with loop-counting and 94\% with sweep-counting. Fig.~\ref{fig:mc_vs_loop}(i) shows traces captured on a 24-core system with the memory-chasing attack for two websites. The traces for different websites are distinct, indicating that the memory-chasing attack captures distinguishable DRAM traffic. In contrast, Fig.~\ref{fig:mc_vs_loop}(ii) illustrates loop-counting traces, which appear indistinguishable, consistent with the low classification accuracy reported in Sec.~\ref{sec:introduction}. 

\textbf{Open-world.}
We observe that memory-chasing still performs marginally better than sweep-counting because memory-chasing traces are more fine-grained than sweep-counting traces. Table~\ref{tab:open-close} summarizes classification accuracies of memory-chasing, sweep-counting, and loop-counting attacks on a 24-core system across both open and closed-world setups.
\clarify{The false positive rate (non-sensitive websites misclassified as sensitive) is 0.16\% and the false negative rate (sensitive websites misclassified as non-sensitive) is 0.22\%.}

\oioioi{\hypertarget{rev:ci}{\textbf{Statistical significance}.}\label{CI} We supplement the key experimental results with standard deviations (SDs) and 95\% confidence intervals (CIs) to quantify the variability and statistical confidence of our measurements. Table~\ref{tab:ci} reports the mean $\pm$ SD across 10 folds for the accuracy metrics, with 95\% $t$-based CIs computed from the fold-level measurements.
For the error rates we report Wilson 95\% CIs. 
The narrow CIs and low fold-to-fold variation indicate that the reported accuracy and error rates are statistically stable across the evaluation samples.}


\oioioi{\textbf{\hypertarget{rev:top}{Effect on top-500 websites.}} We initially evaluate memory-chasing using the 50 most-visited websites from Wikipedia~\cite{siteList}. We now evaluate the memory-chasing attack for a larger closed-world dataset by increasing the number of monitored websites to 500, selected from the Tranco list
~\cite{top500}. \Cref{fig:top500} shows accuracy as the website count increases.}

\begin{table}[t]
\centering
\caption{\oioioi{Statistical confidence of the memory-chasing attack on 24 cores. Results are reported as mean $\pm$ standard deviation (SD) across 10 folds, with 95\% confidence intervals (CIs). Accuracy CIs are $t$-based; $^\dagger$ indicates Wilson~\cite{wilson1927probable} CIs for binomial proportions.}}
\label{tab:ci}
\footnotesize
\oioioi{\begin{tabular}{@{}llcc@{}}
\toprule
\textbf{Setup} & \textbf{Metric} & \textbf{Value (\%)} & \textbf{95\% CI (\%)} \\
\midrule
\multirow{2}{*}{\textbf{Closed-world}}
  & Top-1 accuracy & $96.06 \pm 0.84$ & $[95.46, 96.66]$ \\
  & Top-5 accuracy & $99.30 \pm 0.43$ & $[98.99, 99.61]$ \\
\midrule
\multirow{4}{*}{\textbf{Open-world}}
  & Combined accuracy & $97.19 \pm 0.73$ & $[96.67, 97.71]$ \\
  & Sensitive-site accuracy & $95.86 \pm 1.13$ & $[95.05, 96.67]$ \\
  & False Positive Rate & 0.16 (4/2,500) & $[0.06, 0.41]^\dagger$ \\
  & False Negative Rate & 0.22 (11/5,000) & $[0.12, 0.39]^\dagger$ \\
\bottomrule
\vspace{-0.1in}
\end{tabular}}
\end{table}


The memory-chasing attack outperforms the sweep-counting attack marginally due to the following reasons. First, rather than capturing the effect at LLC, memory-chasing focuses on DRAM, which is also shared but immune to LLC design. Second, memory-chasing counts individual DRAM accesses within each time period rather than counting LLC sweeps. As a result, memory-chasing remains fine-grained regardless of the size of LLC and DRAM. In contrast, larger LLCs cause an increase in time taken to complete one sweep, resulting in fewer sweeps for each time period. This dependence on LLC size causes the sweep-counting attack trace to become coarse-grained, causing a marginal drop in accuracy.

\emph{The loop-counting attack achieves 50\% classification accuracy in the closed-world setup on a 24-core system, remaining low under further constraints like VM isolation and multi-socket systems.} Consequently, we report classification accuracy only for sweep-counting and memory-chasing attacks.

\oioioi{\textbf{\hypertarget{rev:pcMLP}{Effect of memory access patterns and MLP.}} Memory-chasing attack uses random and data-dependent accesses to suppress the effect of hardware prefetching and restrict MLP to one (e.g., via \texttt{lfence}). Table~\ref{tab:benchmarks} isolates these effects on the dual-CCD system. Sequential (non-random) accesses reduce accuracy from 86\% to 72\%, while an increase in MLP reduces it to 69\%. Both changes generate additional attacker DRAM traffic, which introduces self-interference. \emph{In summary, the memory-chasing attack is effective as long as the access pattern uses random addresses and maintains a DRAM MLP of one, minimizing interference and capturing victim-induced DRAM interference.}}


\begin{table}[t]
\centering
\caption{\oioioi{Classification accuracy of memory-chasing attack with different access patterns.}}
\label{tab:benchmarks}
\footnotesize
\oioioi{
\begin{tabular}{@{}lcc@{}}
\toprule
\textbf{Access pattern} &
\textbf{Pointer-chasing} &
\textbf{Parallel} \\
& \textbf{(MLP $=$ 1)} &
\textbf{(MLP $>$ 1)} \\
\midrule
Sequential (Prefetch friendly) & 72 & 47 \\
Random (Prefetch averse) & \textbf{86} & 69 \\
\bottomrule
\vspace{-0.2in}
\end{tabular}
}
\end{table}

\subsubsection{Effect of multiple CCDs}
\label{sec_effect_of_ccds}
We demonstrate that memory-chasing performs well in systems with different LLC designs by evaluating it on an AMD CPU with CCD-based LLC. AMD's chiplet architecture lets us test two scenarios: (i) single-CCD with 16 cores and (ii) dual-CCD with 16 cores split across two CCDs (eight cores per CCD). Table~\ref{tab:isolation_effect} shows the classification accuracy. Memory-chasing has a marginal drop in accuracy from 91\% with the single-CCD configuration to 86\% with the dual-CCD configuration, whereas sweep-counting drops from 84\% to 45\%. We use 16 cores in both scenarios (one and two CCDs) to isolate the impact of CCDs from core counts on classification accuracy. Since memory-chasing relies on DRAM traffic, the LLC decoupling between cores introduced by CCDs primarily impacts sweep-counting. CCDs connect to the I/O die via Global Memory Interconnect (GMI) links, and the I/O die distributes requests to different DRAM controllers. With only one CCD, all DRAM requests originate from a single GMI link~\cite{amd_doc}. For memory-chasing, with two CCDs, the scattering of memory requests across different GMI links causes a slight drop in accuracy. 

Next, we study the effect of both core and CCD counts on the effectiveness of the memory-chasing attack. The classification accuracy is shown in Table~\ref{tab:config_accuracy}. In System B (refer Table~\ref{tab:setup}), each CCD contains two CCXs of eight cores each~\cite{zen4c}. We start with one CCD, both CCXs enabled (16 cores) and evaluate configurations up to four CCDs, increasing core count in steps of 16. As core and CCD counts increase, the accuracy of memory chasing drops. However, it outperforms sweep-counting across all configurations. 

\oioioi{\hypertarget{rev:bumper}As the number of cores and CCDs increases, interrupt processing becomes more distributed across the GMI links and memory controllers. This distribution reduces interference between the attacker and interrupts, thereby degrading classification accuracy. However, accuracy remains at $\sim$72\% even with 64 cores and four CCDs because interrupts remain distributed across all CCDs.}

\begin{table}[t]
\footnotesize
\centering
\caption{Classification accuracy (\%) with memory-chasing and sweep-counting under various isolation mechanisms.}
\label{tab:isolation_effect}
\setlength{\tabcolsep}{3pt}
\begin{tabular}{@{} l cc cc c @{}}
\hline
\multirow{2}{*}{\textbf{Method}} &
\multicolumn{2}{c}{\textbf{CCDs} (16c)} &
\multicolumn{2}{c}{\textbf{Sockets} (24c/s)} &
\textbf{Cross-VM} \\
\cmidrule(lr){2-3}\cmidrule(lr){4-5}
 & \textbf{1} & \textbf{2} & \textbf{1} & \textbf{2} & \textbf{(24c)} \\
\hline
\textbf{Memory-chasing} & \textbf{91} & \textbf{86} & \textbf{96} & \textbf{80} & \textbf{78} \\
Sweep-counting        & 84          & 45          & 94          & 65          & 65 \\
\hline
\end{tabular}
\vspace{-0.1in}
\end{table}

\begin{figure}[t]              
    \centering
    \includegraphics[width=0.7\columnwidth]{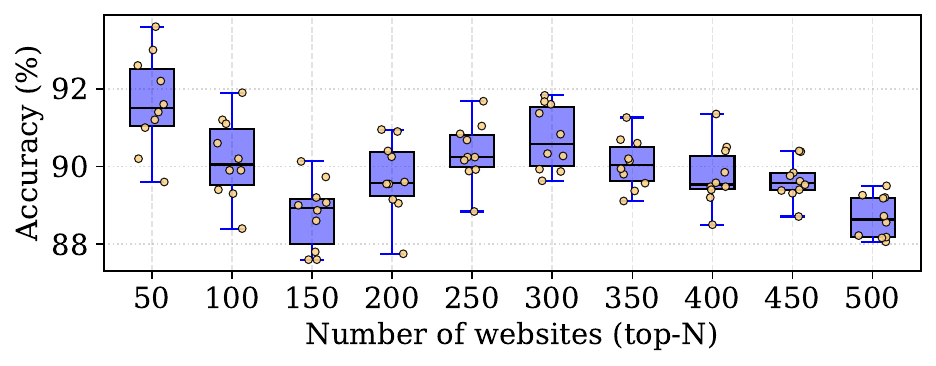}
    \caption{\rebuttal{\oioioi{Classification accuracy of the memory-chasing attack with up to 500 monitored websites from the Tranco list~\cite{top500}.}}}
    \vspace{-0.2in}
    \label{fig:top500}
\end{figure}

\begin{table}[t]
\footnotesize
\centering
\caption{\revision{Classification accuracy (\%) with memory-chasing and sweep-counting for different core and CCD configurations.}}
\label{tab:config_accuracy}
\setlength{\tabcolsep}{4pt}
\begin{tabular}{@{} l cccc @{}}
\hline
\multirow{2}{*}{\textbf{Method}} &
\multicolumn{4}{c}{\textbf{Cores (CCDs)}} \\
\cmidrule(l){2-5}
 & \textbf{16\,(1)} & \textbf{32\,(2)} & \textbf{48\,(3)} & \textbf{64\,(4)} \\
\hline
\textbf{Memory-chasing} & \textbf{91} & \textbf{84} & \textbf{75} & \textbf{72} \\
Sweep-counting        & 84          & 52          & 36          & 26 \\
\hline
\vspace{-0.2in}
\end{tabular}
\end{table}


\subsubsection{Effect of multiple sockets}
In a multi-socket Intel system, each socket has its own LLC shared among its cores. While the LLC is not shared across sockets, cores can access data from other sockets via QuickPath Interconnect (QPI)/Ultra Path Interconnect (UPI) links~\cite{cornellCornellVirtual, frankdennemanNUMADeep, intelmultisocket}, creating NUMA-style performance asymmetry. As a result, sweep-counting accuracy drops from 94\% to 65\% across two sockets (Table~\ref{tab:isolation_effect}). Memory-chasing accuracy also drops from 96\% to 80\% across two sockets, as each socket has its own DRAM controller handling a portion of total DRAM~\cite{stackexchangeRAM}, reducing observable DRAM traffic when the attacker and victim run on different sockets. However, since interrupts are still distributed across sockets, memory-chasing maintains high accuracy.

\simplebox{Takeaway V}{\revision{In multi-socket systems, cross-socket interrupt distribution allows the attacker to observe DRAM traffic without allocating memory on the victim socket.}}

\subsubsection{Effect of VM isolation} \label{socket} 
We evaluate the effect of VM isolation on the memory-chasing attack under a restrictive configuration. 
We deploy two VMs (attacker and victim) on System A (refer Table~\ref{tab:setup}). Table~\ref{tab:isolation_effect} shows classification accuracies of 78\% and 65\% for memory-chasing and sweep-counting, respectively. Memory-chasing outperforms sweep-counting. \revision{QEMU uses \texttt{cgroups} for CPU isolation for VMs~\cite{cgroup}, highlighting that memory-chasing is agnostic of \texttt{cgroup} CPU isolation.} We also isolate the DRAM by placing the attacker and victim VMs on separate sockets. Using \texttt{numactl}~\cite{numactl}, we enforce strict memory allocation locality, reducing classification accuracy approaching a random guess. \oioioi{\hypertarget{rev:numa}{We also evaluate the} effect of different NUMA memory allocation strategies on the accuracy of the memory-chasing attack and observe that \texttt{preferred} and \texttt{membind} policies enforce strict memory allocation locality, reducing classification accuracy approaching a random guess.} 
Since each socket has an independent memory controller handling isolated DRAM, allocating attacker and victim memory to different sockets eliminates DRAM traffic between VMs.

\simplebox{\oioioi{\hypertarget{rev:takeaway}{Takeaway VI}}}{\oioioi{The attack depends on sufficient shared memory-system interference between attacker and victim. Consequently, strict memory-controller/NUMA isolation eliminates the channel.}}

\begin{figure}[t]              
    \centering
    \includegraphics[width=0.7\columnwidth]{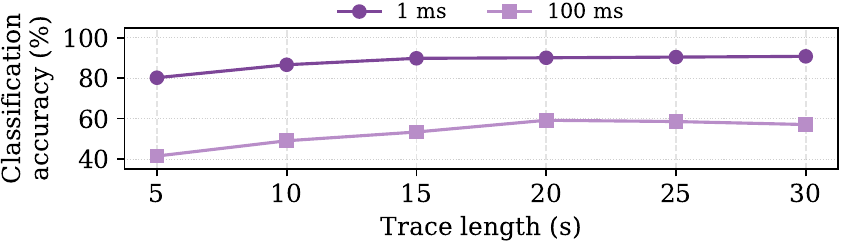}
    \caption{\rebuttal{Effect of trace length and timer resolution on the classification accuracy of the memory-chasing attack for Tor.}}    
    \vspace{-0.2in}
    \label{fig:tor_trace}
\end{figure}

\subsubsection{Effect of different browsers and OSes}
\rebuttal{Classification accuracy remains consistent ($>$95\%) for Chrome and Firefox on both Linux and Windows. We achieve classification accuracy of 91\% for Tor with default settings in contrast to the lower accuracy reported by Cook et al.~\cite{jackcook}. We attribute this primarily to Tor dropping timing precision protections of 100ms~\cite{tor_timer}. Tor floors the default timer precision to match the 60Hz screen refresh rate~\cite{tor_code}. Fig.~\ref{fig:tor_trace} shows the effect of trace length and varying timer resolution on classification accuracy for Tor.}
Finally, across different OSes, Linux uses the Completely Fair Scheduler (CFS)~\cite{linuxsch} while Windows uses preemptive priority-based scheduling~\cite{winsch}. The effectiveness of the attack across both systems indicates that the memory-chasing attack is agnostic to OS scheduling policy.

\subsubsection{Effect of different DRAM configurations}\label{DIMM} Since the memory-chasing attack captures DRAM traffic, we evaluate it under different DIMM counts, DRAM speeds, and interleavings. For DIMM count, we collect traces with one and eight DIMMs, and observe classification accuracy consistently above 90\%. For different DRAM speed, the trace shifts downwards with decreasing DRAM speed; however, the normalized trace remains the same, and accuracy is unchanged. For interleaving, we use Intel and AMD DRAM controllers, each employing a different hash function, and observe high accuracies (Intel: 96\%, AMD: 91\%).
\begin{table}[t]
\footnotesize
\centering
\caption{Classification accuracy (\%) with memory-chasing and sweep-counting on 24 cores under countermeasures.}
\label{tab:accuracy_cm}
\setlength{\tabcolsep}{4pt}
\begin{tabular}{@{} l c c c c @{}}
\hline
\multirow{2}{*}{\textbf{Method}} &
\multirow{2}{*}{\textbf{Baseline}} &
\multicolumn{3}{c}{\textbf{Countermeasures}} \\
\cmidrule(l){3-5}
 & & \textbf{CS}~\cite{shusterman2019robust} & \textbf{SI}~\cite{jackcook} & \textbf{DNI} \\
\hline
\textbf{Memory-chasing} & \textbf{96} & \textbf{91} & \textbf{74} & \textbf{81} \\
Sweep-counting        & 94          & 75          & 62          & 76          \\
\hline
\end{tabular}
\smallskip

\footnotesize CS\,=\,Cache-sweep noise, SI\,=\,Spurious-interrupt, DNI\,=\,DRAM-noise injection
\vspace{-0.2in}
\end{table}
\subsubsection{Effect of interrupt scheduling and shaping} \mbox{} \\ \label{interrupt_isolation}
\oioioi{\textbf{Interrupt scheduling.} Linux provides \texttt{irqbalance} to control interrupt distribution across cores. We evaluate the effect of interrupt scheduling by using affinity tuning to isolate the interrupt to different cores under three configurations: (i) \textit{baseline}: \texttt{irqbalance} enabled with the default policy; (ii) \textit{isolated attacker}: interrupts disabled on the attacker core; and (iii) \textit{isolated interrupts}: all interrupts pinned to a single non-attacker core. Traces for the memory-chasing attack remain similar in all cases, with a correlation value $>0.95$, indicating that it is agnostic to interrupt isolation and scheduling.} 

\oioioi{\textbf{Interrupt shaping.} We evaluate Linux mechanisms that regulate interrupt rate or timing, one at a time: fixed NIC interrupt coalescing (\texttt{ethtool -C}), NAPI interrupt deferral (\texttt{napi\_defer\_hard\_irqs}), NAPI work bounding (\texttt{netdev\_budget}), threaded NAPI, \texttt{tc} traffic shaping, and timer slack. Traces for the memory-chasing attack remain highly correlated in all cases across all websites, with a correlation value $>0.89$, indicating that interrupt shaping does not impact the memory-chasing attack. We also discuss the implications of extending DRAM traffic shaping solutions to include interrupts in \Cref{sec:countermeasure}.} 



\subsubsection{Effect of system noise} We evaluate the effect of noisy environments on classification accuracy. We define two classes of noise: class-I and class-II. For class-I noise, we open 10 different websites on Firefox, with two co-running applications, Visual Studio Code and Slack. Before loading the victim website on a different browser, Chrome, we load two additional websites, \texttt{apple.com} and \texttt{spotify.com}. For class-II noise, we stream high-resolution videos (up to 4K) on YouTube. We also concurrently play a video in a loop, and run \texttt{stress-ng}, stressing CPU and memory.
Our attack achieves accuracies of 94\% and $\sim$80\% with class-I and class-II noise, respectively. The key insight explaining this robustness is that, in class-I noise, background tabs and idle processes stay in a steady state, generating negligible interrupt activity, and thus do not interfere. As mentioned by Cook et al.~\cite{jackcook}, loop-counting also remains unaffected by co-running applications. With class-II noise, we observe a slight drop in accuracy, attributed to interference from additional network interrupts.

\subsection{Privacy and security implications}\label{subsec:process_finger_eval}
Beyond WF, memory-chasing can be adapted to a wide range of side-channel attacks. We demonstrate two additional exploits that underscore the attack's generality.

\textbf{Process Fingerprinting.} We extend the memory-chasing attack to identify processes on the victim's system. The attacker, as a malicious website, collects traces to predict the applications the victim launches. The threat model mirrors the WF attack, except the adversary targets processes rather than websites. 
Memory-chasing achieves 98\% and 93\% classification accuracy in single-socket (24-core) and dual-socket (48-core) configurations, respectively; while sweep-counting drops from 93\% to 67\% across the same configurations. 

\textbf{Inter-keystroke timing.} Memory-chasing can infer keystroke activity. The attacker records a memory-chasing trace while the victim types.
Dips in the trace coincide with the victim's keypresses, indicating that the attacker can detect keystrokes. \rebuttal{Over a 10-second trace, the attack produces approximately four false positives. With a 5ms interval, the timing error, measured as mean absolute error between the keystroke and the attacker's observation, is 2.48ms.}

\section{Countermeasures} \label{sec:countermeasure}
We evaluate the effect of countermeasures based on prior works~\cite{jackcook} ~\cite{shusterman2019robust}, on the classification accuracy of the memory-chasing attack. We consider: (i) cache-sweep noise~\cite{shusterman2019robust}, which repeatedly evicts cache lines to disrupt LLC-occupancy attacks like sweep-counting. Analogously, since memory-chasing operates at DRAM, we propose and evaluate (ii) DRAM noise injection, which introduces artificial bandwidth contention to mask the victim’s memory traffic. For DRAM noise injection, we use \texttt{stress-ng}~\cite{stressng} with the \texttt{--brk} flag, limited to 80\% of DRAM capacity. This flag causes \texttt{stress-ng} to repeatedly allocate and access 4KB pages, avoiding lazy-loading of the pages. We also consider (iii) timer randomization~\cite{jackcook}, which randomizes the browser timer to disrupt attacks relying on precise timing measurements, and (iv) spurious interrupt injection~\cite{jackcook}, which injects artificial interrupts to add noise to interrupt-based measurements. Table~\ref{tab:accuracy_cm} summarizes their impact on memory-chasing and sweep-counting attacks. 
As mentioned in \Cref{sec:eval}, the accuracy of the loop-counting attack in many-core systems is itself low ($\sim50\%$), so we do not evaluate it under countermeasures.

\textbf{Cache-sweep noise and DRAM noise injection.} In 24 cores, with cache-sweep noise, we observe a drop in accuracy from 96\% to 91\% for memory-chasing and from 94\% to 75\% for sweep-counting. For DRAM noise injection, memory-chasing classification accuracy is 81\%, outperforming sweep-counting, whose accuracy is 76\%.


\oioioi{\textbf{\hypertarget{rev:timer}{Timer randomization}.} \label{timer} The classification accuracy drops from 96\% without timer randomization to 26\% with a randomized timer in Chrome for memory-chasing attack. Timer randomization has not yet been deployed in any real world browser because it precludes websites that require a high-resolution timer, as mentioned by Cook et al.~\cite{jackcook}. Nevertheless, to circumvent the dependence on the browser timer, we \emph{test} a custom-made timer to mount the memory-chasing attack. We use a JS thread with a shared variable to create a timer, previously explored by Schwarz et al.~\cite{JStimergruss2017}. The attacker creates two JS threads and shares a variable across the two threads. Thread one increments the shared counter, and thread two uses the value of the shared counter as a timer to perform the memory-chasing attack.
Using a timer-free attack, we bypass timer randomization and record a classification accuracy of 90\% when tested on Chrome. \emph{However, the practicality of timer free attacks is constrained by the strict shared memory requirements as mentioned in \Cref{subsec:multithread}. Additionally, Tor browser completely prohibits shared variables between threads, rendering the timer-free attack infeasible on Tor.
} }

\textbf{Spurious-interrupt noise.} The accuracy of the memory-chasing attack drops to 74\% from 96\% with spurious interrupts. However, it still outperforms the sweep-counting attack, which drops to 62\% from 94\%. \clarify{Spurious interrupt injection is a viable countermeasure against the memory-chasing attack, which relies on interrupts as its side channel, but incurs a $\sim$16\% website load-time overhead~\cite{jackcook}.}

\textbf{Shared resource QOS mechanisms.}
Shared Resource QOS mechanisms operate at shared resources like LLC and DRAM bandwidth. Intel RDT~\cite{IntelRDT} regulates LLC and memory bandwidth across workloads, as does AMD PQOS~\cite{amdqos}. \rebuttal{We evaluate under two allocations of DRAM bandwidth: (i) attacker 70\% and victim 10\%, and (ii) attacker 10\% and victim 70\%.} Classification accuracy remains above 93\% for Intel RDT and above 90\% for AMD PQOS. Since these mechanisms do not partition interrupts, interrupt activity remains observable~\cite{jackcook}.
LLC partitioning techniques isolate the LLC across security domains~\cite{intelCAT, kiriansky2018dawg, costan2016sanctum, bourgeat2019mi6, saileshwar2021bce}, eliminating inter-domain contention at the LLC. LLC partitioning does not defend against the memory-chasing attack, which neither creates LLC interference nor forces the victim’s accesses to DRAM. 


\oioioi{\textbf{\hypertarget{rev:oram}{DRAM traffic shaping and ORAM.}}} DRAM traffic shaping techniques eliminate timing side channels by shaping memory access patterns, either through deterministic scheduling or statistical traffic normalization. Defenses such as DAGguise~\cite{deutsch2022dagguise} and Camouflage~\cite{zhou2017camouflage} allow dynamic traffic contention to maintain performance. As these defenses~\cite{zhou2017camouflage, deutsch2022dagguise} preserve dynamic DRAM contention and do not partition interrupts, the memory-chasing attack remains effective. \oioioi{Extending these defenses to interrupt traffic would require either (i) shaping DRAM traffic that varies with a webpage load, which neither scales to a large set of websites nor hides the visited site, since the tailored shape itself differs per website, or (ii) enforcing one worst-case shape for all websites, i.e., a static DRAM service. Only the latter removes the channel, so effective mitigation ultimately requires a system-wide, fixed-rate static DRAM service~\cite{FixedService} across all websites, incurring a prohibitive performance overhead. An alternate approach is Oblivious RAM (ORAM)~\cite{stefanov2013pathoram, haider2017flat, ren2013design}, which hides \emph{which} memory addresses a program accesses. The memory-chasing attacker does not observe addresses; instead, it measures aggregate contention at the memory controller. Thus, ORAM also cannot prevent the memory-chasing attack.}

\oioioi{\textbf{\hypertarget{rev:secandperf}{Security and performance} tradeoff of countermeasures.} The evaluated countermeasures so far degrade attack classification accuracy at a performance or functionality cost. Cache-sweep and DRAM noise leave memory-chasing accuracy above 80\% while consuming CPU time and memory bandwidth. Spurious interrupts degrade accuracy to 74\% but add $\sim$16\% page-load overhead~\cite{jackcook}. To our knowledge, timer randomization degrades accuracy to 26\% but breaks websites that require high-resolution timers and is not deployed in major browsers. Shared-resource QoS leaves accuracy above 90\%, while LLC partitioning and DRAM traffic shaping do not eliminate the signal. Static DRAM service incurs prohibitive overhead~\cite{FixedService}. Thus, the countermeasures impose substantial performance or functionality costs, limiting their practicality.}


\section{Robustness}
We demonstrate the robustness of the memory-chasing attack across multiple dimensions. \rebuttal{We assume the availability of the default timer exposed by browsers.} Memory-chasing scales to systems with many cores, whereas the prior state-of-the-art loop-counting loses effectiveness, as shown in \Cref{sec:loop_increase_core_count}. Our attack is insensitive to LLC design and remains effective regardless of CCD-based designs, whereas sweep-counting loses effectiveness. Memory-chasing operates reliably on a system with multiple sockets and under VM isolation. While spurious interrupt injection could mitigate the attack, browsers do not employ it in practice due to its negative impact on performance, power efficiency, and security guarantees.

\textbf{Effect of huge pages.} \clarify{The memory-chasing attack is invariant to Transparent Huge Page (THP) configuration; 2MiB pages shift counter values higher due to more TLB hits, but the trace pattern and classification accuracy remain unchanged.}

\textbf{Effect of network conditions.} \rebuttal{Network speed affects website load time and impacts classification accuracy when the model trains on traces recorded at vastly different speeds, e.g. 100 kbps versus 5–10 Mbps. However, if traces are recorded at typical browsing speeds of 5–10 Mbps, the accuracy remains unchanged. Assuming average network conditions, classification relies solely on how the website loads on the victim's end, which the memory-chasing attack captures. To verify, we limit network speed using \texttt{wondershaper}~\cite{wondershaper_github} and collect traces at different speeds. We find that above 4–5 Mbps, traces are consistent, and classification accuracy is unchanged when the test traces are recorded at different speeds.}

\textbf{Effect of login state and user personalization.} \rebuttal{We evaluate the effect of logged-in and login landing pages across three representative websites: \texttt{youtube.com}, \texttt{x.com}, and \texttt{reddit.com}. Traces vary depending on whether the website serves personalized content or an unauthenticated landing page. However, traces across different user profiles remain similar, indicating that user-specific recommendations have little effect on the traces. Similarly, for websites with location-specific recommendations, the attack remains invariant to location changes. To handle logged-in and logged-out pages, we train on traces from both states and assign each a separate class, achieving high classification accuracy.}

Collectively, these results establish memory-chasing as a robust and scalable WF attack.

\section{Related Work}
\textbf{Network-based WF attacks.}
Network attacks exploit traffic characteristics. Early work~\cite{hintz2002fingerprinting} revealed packet size and timing leakage; subsequent attacks targeted Tor~\cite{cherubin2022evaluating}~\cite{aminuddin2023rise} using machine learning~\cite{herrmann2009website, pachenko2011website, cai2012touching, wang2013improved, gong2012website}. Scalable attacks~\cite{panchenko2016website}~\cite{hayes2016k} and deep learning~\cite{rimmer2018automate}~\cite{sirinam2018deep} improved accuracy despite challenges~\cite{juarez2014critical}. 
Defenses employ traffic padding and shaping~\cite{cai2014systematic, cai2014csbuflo, nithyanand2014glove, wang2017walkie, Cherubin2017WebsiteFD, mcguan2024practical}, 
but fail against microarchitectural channels. Memory-chasing observes DRAM traffic rather than network traffic, bypassing all network defenses.

\textbf{Other microarchitectural side channels.} 
Diverse attacks exploit specialized hardware. GPU vulnerabilities leak webpage content~\cite{lee2014stealing} but were patched. Performance counters enable fingerprinting~\cite{gulmezoglu2017perfweb} but require privileged access. Memory-chasing attack requires neither special hardware nor privileges. JS event loops~\cite{vila2017loophole} and memory footprints~\cite{jana2012memento} provide coarse leakage with low accuracy. Other channels include storage patterns~\cite{kim2016inferring}, Quick UDP Internet Connections (QUIC) protocol~\cite{siby2023evaluating}, interrupt timing~\cite{rauscher2024crosscore}, and CPU idle states~\cite{rauscher2024idleleak}.  

\section{Conclusion}\label{sec:conclusion}
We proposed a memory-chasing attack that leverages DRAM traffic as a scalable and robust side channel for WF. In many-core systems, we showed that loop-counting and sweep-counting attacks become less effective. We argued that a system-wide shared resource, such as DRAM, constitutes an attractive attack surface. \revision{We demonstrated the correlation between interrupts and DRAM traffic.}
We serialized DRAM accesses via pointer chasing, ensuring at most one in-flight DRAM request, minimizing interference while preserving sensitivity to interrupt-driven DRAM activity. 
Our proposed attack captured DRAM traffic caused by interrupts, and achieved 96\% classification accuracy for the top-50 most visited websites on 24-core systems. Our intensive evaluation spanning many-core configurations, LLC designs, browsers, operating systems, sockets, VMs, \revision{system noise}, and countermeasures, demonstrated the robustness of the memory-chasing attack. 


\balance
\bibliographystyle{IEEEtranS}
\bibliography{sample-base}

\end{document}